\documentclass{edm_article}
\usepackage{booktabs}
\usepackage{multirow} 
\usepackage{graphicx} 

\begin{document}

\title{How Good Are Large Language Models for Course Recommendation in MOOCs?}

\numberofauthors{5}
\author{
\alignauthor
Boxuan Ma\\
\affaddr{Kyushu University, Japan}\\
\email{boxuan@artsci.kyushu-u.ac.jp}
\alignauthor 
Md Akib Zabed	Khan\\
        \affaddr{Florida International University, United States}\\
        \affaddr{United States}\\
        \email{mkhan149@fiu.edu}
\alignauthor Tianyuan	Yang\\
\affaddr{Kyushu University, Japan}\\
\email{yangtianyuan1108@gmail.com}
\and    
 \alignauthor Agoritsa	Polyzou\\
      \affaddr{Florida International University, United States}\\
        \email{apolyzou@fiu.edu}
 \alignauthor Shin'ichi	Konomi\\
\affaddr{Kyushu University, Japan}\\
\email{konomi@artsci.kyushu-u.ac.jp}
}

\maketitle

\begin{abstract}
Large Language Models (LLMs) have made significant strides in natural language processing and are increasingly being integrated into recommendation systems. However, their potential in educational recommendation systems has yet to be fully explored. This paper investigates the use of LLMs as a general-purpose recommendation model, leveraging their vast knowledge derived from large-scale corpora for course recommendation tasks. We explore a variety of approaches, ranging from prompt-based methods to more advanced fine-tuning techniques, and compare their performance against traditional recommendation models. Extensive experiments were conducted on a real-world MOOC dataset, evaluating using LLMs as course recommendation systems across key dimensions such as accuracy, diversity, and novelty. Our results demonstrate that LLMs can achieve good performance comparable to traditional models, highlighting their potential to enhance educational recommendation systems. These findings pave the way for further exploration and development of LLM-based approaches in the context of educational recommendations.

\end{abstract}

\keywords{Course Recommendation, Large Language Models, Recommendation System} 

\section{Introduction}

Course recommendation systems are increasingly used in the field of education and have become an essential tool in addressing information overload and enhancing user experience for learning \cite{ma2020course}. These systems can offer personalized course suggestions that align with a student's interests, career goals, or skill development needs. This personalized approach can make learning environments more adaptive and effective, enhancing the educational experience by helping students navigate the vast array of available courses and make more informed decisions about their learning journey \cite{jiang2019goal}.

Over the past decade, significant advancements have been made in course recommendation technologies, particularly with the rise of deep learning and large-scale data-driven models \cite{ma2024survey}. Traditional recommendation models, including collaborative filtering and content-based methods, have long been employed in practical settings. These approaches typically rely on user-item interaction data or explicit features to provide personalized recommendations \cite{ma2021investigating}. While successful, traditional recommendation models face notable limitations, such as a lack of generalization and the need for task-specific data for training. On the other hand, deep learning models have demonstrated considerable potential in enhancing prediction accuracy, but they require extensive training and often suffer from a lack of explainability, making them less transparent to users \cite{dai2023uncovering}. 

In recent years, Large Language Models (LLMs), such as ChatGPT, have gained significant attention due to their great performance in a variety of natural language processing tasks, including text generation, question answering, and language comprehension \cite{memarian2023chatgpt}. These models, with their adaptability and vast knowledge derived from large-scale corpora, present an appealing opportunity for recommendation systems. Previous research has indicated LLMs can be directly used as recommendation systems with prompts, and a growing body of research has begun to explore the potential of LLMs in recommendation tasks \cite{di2023evaluating}. Several studies have explored the potential of LLMs as zero-shot recommendation systems, evaluating their performance across various recommendation scenarios and datasets from different domains \cite{di2023evaluating,dai2023uncovering,liu2023chatgpt}. Their results show that LLMs are capable of adapting to different recommendation scenarios and improving system performance without the need for task-specific training data. On the other hand, many researchers have started using LLMs as part of recommendation systems to enhance their performance, such as through feature extraction, feature augmentation, or knowledge representation. In the educational domain, Yang et al. use LLMs to generate knowledge concepts from course descriptions \cite{yang2024leveraging} and provide course recommendations based on LLM-generated concepts \cite{tianyuan2024boosting}.

Despite the rapid development of LLMs, most research on LLM-based recommendation systems has focused on domains like music, movies, and books. There has been limited research on applying LLMs specifically to course recommendations within the context of Massive Open Online Courses (MOOCs), and whether LLMs can perform well on course recommendation tasks remains an open question. Therefore, this paper aims to bridge this gap by exploring the potential of LLMs for course recommendation in MOOCs. We evaluate the effectiveness of LLMs in recommending courses based on user learning history. Our study offers a comparative analysis between LLMs and traditional recommendation models and investigates the promise of LLMs in addressing key challenges in educational recommendation systems.

\begin{figure*}[htb]
\centering
\includegraphics[scale=0.677]{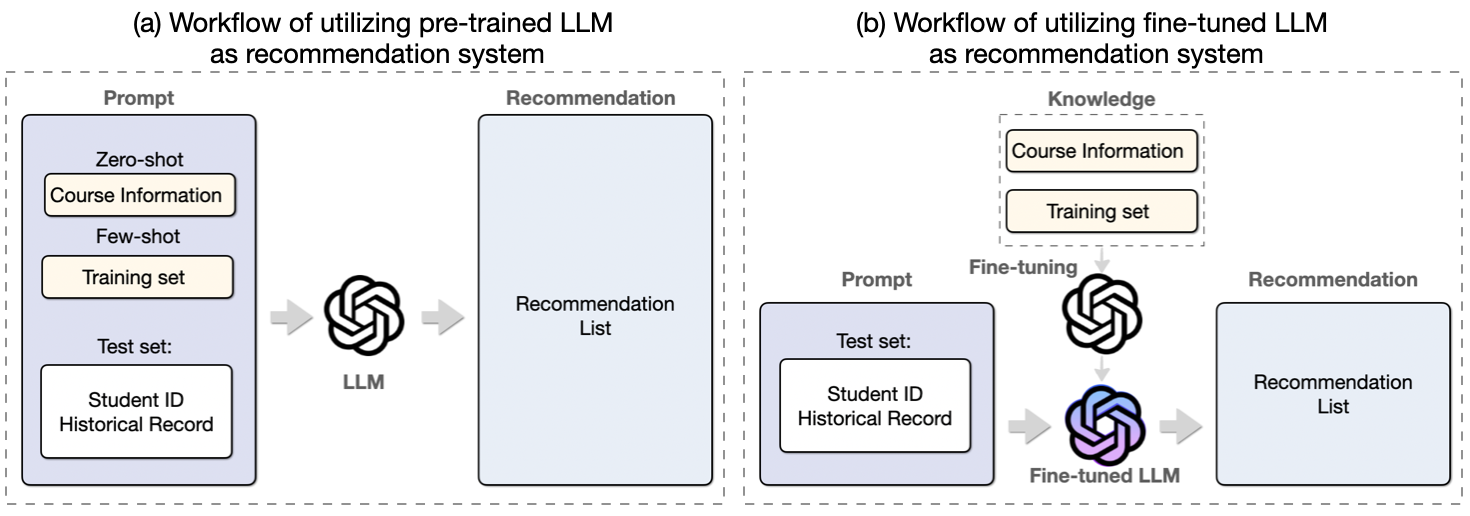}
\caption{Workflow of utilizing LLMs to perform course recommendation tasks.}
\end{figure*}

\section{Related Work}
\subsection{Course Recommendation}
Recommending courses to students is a critical yet challenging task, as their course choices can influence future learning paths, skill development, and career decisions \cite{ma2021investigating}. The rise of Massive Open Online Courses (MOOCs) and the increasing number of students have led to the widespread application of course recommendation systems. 

Since the introduction of the first course recommendation system based on constraint satisfaction \cite{parameswaran2011recommendation}, various methods have been developed. Content-based approaches recommend courses by matching students’ interests with course descriptions and content \cite{jing2017guess, morsomme2019content, morsy_sara_2019_3554678, naren2020recommendation}. Matrix Factorization (MF) techniques have also been applied to course recommendation, particularly for predicting future course selections based on students’ past courses and grades \cite{elbadrawy2015collaborative, sweeney2016next}. Other methods have explored the mining of historical course enrollment data to uncover relationships and patterns. For example, \cite{aher2013combination} employed association rule mining combined with clustering to identify course relationships, recommending courses based on historical enrollment patterns. Similarly, \cite{bendakir2006using} used association rules with user ratings to enhance the recommendation results, while \cite{polyzou2019scholars} introduced Scholars Walk, which captures sequential course relationships through a random-walk approach.

As deep learning techniques have gained popularity, they have also been applied to course recommendation systems \cite{gong2020attentional, zhang2019hierarchical, pardos2019combating, pardos2019connectionist, jiang2019goal}. For instance, \cite{pardos2019combating} modified the skip-gram model to generate course vectors from historical course enrollment data, which are then used to recommend courses similar to a student's previously favored courses. In a similar vein, \cite{pardos2019connectionist} proposed the course2vec model, which employs a neural network to generate course recommendations by taking multiple courses as input and predicting a probability distribution over potential course selections.

While traditional models have significantly advanced recommendation performance, they often suffer from requiring extensive training. In addition, their black-box nature often complicates model interpretability \cite{ma2021exploration, ma2021courseq}.

\subsection{LLMs for Recommendation}

Large Language Models (LLMs) have demonstrated their adaptability and significant improvements in a wide range of natural language processing (NLP) tasks by leveraging the extensive knowledge from large-scale corpora. Inspired by its successes, there has been a growing interest in applying LLMs to recommendation systems. 

A number of recent works have leveraged prompt-based techniques to transform recommendation tasks into natural language tasks, utilizing LLMs without task-specific fine-tuning. For instance, LMRecSys \cite{zhang2021language} and P5 (Pretrain, Personalized Prompt, and Predict Paradigm) \cite{geng2022recommendation} focus on converting recommendation tasks into multi-token cloze tasks using prompts to tackle zero-shot and data efficiency issues. GPT4Rec \cite{li2023gpt4rec} and M6-Rec \cite{cui2022m6} utilize LLMs to learn both item and user embeddings. Liu et al. \cite{liu2023chatgpt} evaluated ChatGPT’s performance on five recommendation scenarios, including rating prediction, sequential recommendation, direct recommendation, explanation generation, and review summarization. Dai et al. \cite{dai2023uncovering} investigated ChatGPT’s ranking capabilities, including point-wise, pair-wise,  and list-wise ranking. Moreover, the ability of LLMs has also been explored in cold-start scenarios where few user interaction data are available \cite{wang2023zero,wu2023towards}.

Besides the direct use of LLMs as recommendation systems, LLMs are increasingly being used as components to enhance traditional recommendation models. These approaches integrate LLMs into existing systems through feature extraction, feature augmentation, knowledge representation, and ranking functions \cite{wu2024survey}. For instance, Gao et al. \cite{gao2023chat} are among the first to use ChatGPT to augment traditional recommender systems by injecting user preferences into the recommendation process through conversational interaction. Another example, Zhang et al. \cite{zhang2023recommendation} enhance recommendation system with LLMs by designing prompts for different recommendation settings, where LLM takes candidates from a Recall model for re-ranking.

Despite the growing body of research on LLM-based recommendation systems in domains like music, movies, and books, there has been limited exploration of LLMs for educational settings, specifically for course recommendations. To our knowledge, there has been limited research on applying LLMs specifically to course recommendations aside from work by Khan et al. \cite{khan2022can}. However, their work did not focus on the evaluation of LLMs' potential and only used local models, and whether LLMs can perform well on course recommendation tasks remains an open question. Given the promising results from LLM applications in other domains, we conduct a thorough evaluation of their capabilities in course recommendation tasks.

\begin{figure}[ht]
\centering
\includegraphics[scale=0.51]{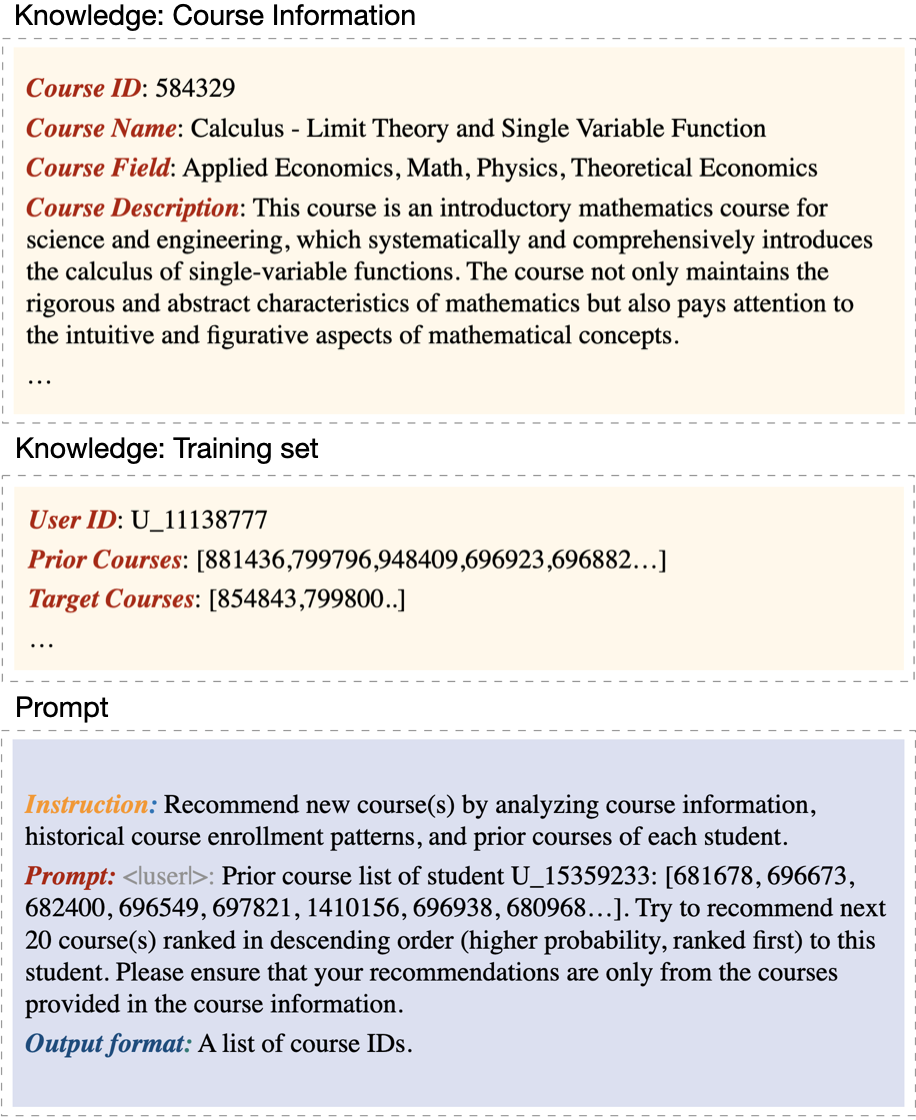}
\caption{Example prompt of course recommendation task.}
\end{figure}

\begin{figure}[ht]
\centering
\includegraphics[scale=0.525]{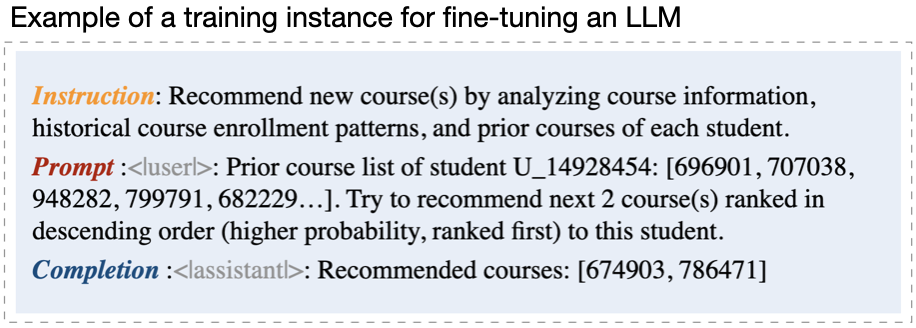}
\caption{Example of a training instance for fine-tuning.}
\end{figure}

\section{Methods} 

The workflow for using LLMs in course recommendation tasks is shown in Figure 1. We explore two approaches for applying LLMs to course recommendations. The first approach involves using pre-trained LLMs as recommendation systems, where they generate recommendations directly based on prompts. The second approach involves fine-tuning the model, enriching its knowledge base with student interaction data, and generating recommendations based on the fine-tuned model.

\subsection{Direct Use of LLMs for Recommendation}  

In this approach, we directly use LLMs to generate recommendations without re-training or fine-tuning the model with training data. Instead, we craft prompts and feed them into the LLMs. The model then generates recommendation results based on the specific instructions provided in the prompts.

For zero-shot learning, we provide the LLM with information about all available courses (including course IDs, names, and descriptions), along with the student's prior course registration history as input. The LLM is tasked with recommending a set of courses based on this input. In the case of few-shot learning, we incorporate additional training data as context and prompt the LLM to recommend courses based on the complete set of provided information. An example of course information, training set, and prompt can be seen in Figure 2.  

\subsection{Fine-tuning LLMs for Recommendation}  

We also fine-tune LLMs to enhance their knowledge with historical data relevant to our task. We fine-tune two open-source models, Llama-3 \cite{touvron2023llama} and GPT-2 \cite{radford2019language}, as they are freely available and easy to use. Following prior work in item recommendation \cite{liu2023llmrec}, we use students' course enrollment histories to fine-tune the LLMs, enabling them to capture historical enrollment patterns. After fine-tuning, we provide prompts that include a student's prior course registration history and ask the fine-tuned models to recommend a set of courses based on this information.

An example of a preprocessed training instance is shown in Figure 3. We use <|user|> and <|assistant|> tokens to indicate the input and output in each training instance, where the input is the student's prior course list, and the output is the subsequent courses they are likely to take. Additionally, we include course descriptions as input to help the model capture the semantic similarity between courses.

\section{Evaluation}

\subsection{Dataset} 
In the context of this work, we focus on the scenario of course recommendation within a MOOC environment. The dataset MOOCCubeX \cite{yu2021mooccubex} in our analysis is collected from the XuetangX\footnote{http://www.xuetangx.com}, one of the largest MOOC websites in China. This publicly available dataset consists of 4,216 courses and more than 3,330,294 students. We preprocessed the data to better model users and courses by filtering out users with few interactions. Specifically, we retained users with more than five interactions and courses with more than ten interactions in our work.

\subsection{Baselines}

We explore different ways to use LLMs for the task of course recommendation, from direct prompt-based methods to fine-tuning strategies that enrich the knowledge base with student interaction data. For the direct recommendations LLMs, we use \textbf{GPT4-turbo} and \textbf{GPT4o} because of their popularity and affordability. Following previous work \cite{khan2022can}, we fine-tuned two open-source models, \textbf{Llama-3} \cite{touvron2023llama} and \textbf{GPT2} \cite{radford2019language}.

We also compared LLMs to following traditional baselines: \textbf{Random}, recommend random items for users. \textbf{Pop} \cite{elbadrawy2016domain}, provides the most popular items for users. \textbf{PMF} \cite{salakhutdinov2007probabilistic}, a traditional recommendation model that relies solely on the user-item rating matrix. \textbf{NMF} \cite{Lee1999Learning}, factorizes a non-negative matrix into the product of two or more non-negative matrices based on the user-item interaction matrix. \textbf{Item-based KNN} \cite{sarwar2001item}, models user and item based on item similarity obtained by interaction information. \textbf{User-based KNN} \cite{sarwar2001item}, models user and item based on user similarity obtained by interaction information.

\subsection{Evaluation Metrics}
To get a comprehensive evaluation that sheds light on LLMs' performance in the course recommendation task, we select the different metrics following previous works \cite{dai2023uncovering,di2023evaluating,liu2023chatgpt}. 

For accuracy metrics, we utilize various metrics, including \textbf{Hit Ratio}, \textbf{Recall}, \textbf{Precision}, and \textbf{F1}. Furthermore, we employ the normalized Discounted Cumulative Gain (\textbf{nDCG}) metric to evaluate the quality of the ranking in the recommendation list. Higher scores in these metrics indicate better recommendations. 

For coverage and novelty metrics, we select \textbf{ItemCoverage}, \textbf{Gini Index} and Expected Popularity Complement (\textbf{EPC}). ItemCoverage quantifies the coverage of all available items that can potentially be recommended, and the Gini Index assesses the distribution of items. These two metrics measure the diversity of recommendations. Additionally, EPC measures the expected number of relevant recommended items that were not previously seen by the user, showing the model's ability to introduce novelty in the recommendations. Higher scores in these metrics indicate better recommendations.

\begin{table*}[t]
    \centering
    \caption{Accuracy performance comparison $(\%)$}
    \resizebox{\textwidth}{!}{
    \begin{tabular}{lccccccccccccc} \toprule
        & \multicolumn{5}{c}{K = 5} & \multicolumn{5}{c}{K = 10} \\
        \cmidrule(lr){2-6} \cmidrule(lr){7-11}
        Model & Hit Ratio@5 $\uparrow$ & Recall@5 $\uparrow$ & Precision@5 $\uparrow$ & F1@5 $\uparrow$ & nDCG@5 $\uparrow$ & Hit Ratio@10 $\uparrow$& Recall@10 $\uparrow$& Precision@10 $\uparrow$& F1@10 $\uparrow$& nDCG@10 $\uparrow$\\
        \midrule
        
        Random & 0.100 & 0.005 & 0.020 & 0.010 & 0.020 &0.610&0.009 & 0.080 & 0.090 & 0.080\\
        
        GPT4-turbo zero-shot &0.210&0.100&0.040&0.060&0.050&0.410 & 0.310 & 0.040 & 0.070 & 0.130 \\
        GPT4o zero-shot & 0.405 & 0.080 & 0.080 & 0.075 & 0.075 & 0.815 & 0.185 & 0.080 & 0.110 & 0.110 \\
              Item-based KNN & 0.510 &0.070& 0.100&0.080&0.125&1.225 &0.400&0.130&0.195&0.215\\
              GPT4o few-shot  &0.800&0.400&0.160&	0.230&0.230& 0.800&0.400&0.080&	0.130&0.230 \\
        GPT4-turbo few-shot  &1.002&0.113&0.201&0.147&0.015&1.000&0.110&	0.100&0.110&	0.100& \\

       PMF & 1.630 & 0.285 & 0.325 & 0.280 & 0.435 & 3.370 & 0.680 & 0.340 & 0.425 & 0.515 \\
        NMF & 1.630 & 0.285 & 0.325 & 0.280 & 0.435 & 3.370 & 0.680 & 0.340 & 0.425 & 0.515 \\
        User-based KNN & 2.960 &1.080&0.595&0.755&0.955 &4.595&1.645&0.460&0.715&1.100\\
       Pop & 8.195 & 3.300 & 1.680 & 2.215 & 2.680 & 15.165 & 5.950 & 1.660 & 2.580 & 3.640 \\
GPT2 Fine-tuning & \underline{16.903} & \underline{7.438} & \underline{3.524} & \underline{3.855} & \underline{11.492} & \underline{22.643} & \underline{9.560} & \underline{2.452} & \underline{3.483} & \underline{13.498} \\
Llama3 Fine-tuning & \textbf{21.677} & \textbf{12.434} & \textbf{4.852} & \textbf{5.939} & \textbf{15.266} & \textbf{28.857} & \textbf{15.166} & \textbf{3.424} & \textbf{4.770} & \textbf{17.496} \\

        \midrule
        & \multicolumn{5}{c}{K = 15} & \multicolumn{5}{c}{K = 20} \\
        \cmidrule(lr){2-6} \cmidrule(lr){7-11}
        Model & Hit Ratio@15 $\uparrow$ & Recall@15 $\uparrow$ & Precision@15 $\uparrow$ & F1@15 $\uparrow$ & nDCG@15 $\uparrow$ & Hit Ratio@20 $\uparrow$ & Recall@20  $\uparrow$ & Precision@20 $\uparrow$ & F1@20 $\uparrow$ & nDCG@20 $\uparrow$ \\
        \midrule
        Random &1.230&0.150 & 0.090 & 0.110 & 0.110 & 1.435 & 0.190 & 0.070 & 0.100 & 0.115 \\
           GPT4-turbo zero-shot&0.410&0.310&0.030&0.050& 0.130 & 1.230 & 0.940 & 0.060 & 0.120 & 0.280 \\
              GPT4o zero-shot &1.225 & 0.450 & 0.085 & 0.135 & 0.185 & 1.225 & 0.450 & 0.060 & 0.105 & 0.185 \\
                Item-based KNN&2.450&0.590&0.175&0.270&0.320&3.165&0.845& 0.175 & 0.295 & 0.385 \\

      GPT4o few-shot  &0.800&0.400&0.500&0.900&0.230&0.800&0.400&	0.400&0.700&0.230  \\
             GPT4-turbo few-shot&1.000&0.110&0.070&0.080&0.100&3.000&0.118&0.150&0.270&0.370\\
    
        PMF & 5.205 & 1.465 & 0.345 & 0.560 & 0.885 & 5.100 & 0.980 & 0.260 & 0.405 & 0.620 \\
        NMF & 5.205 & 1.465 & 0.345 & 0.560 & 0.885 & 5.100 & 0.980 & 0.260 & 0.405 & 0.620 \\
      User-based KNN&6.530&2.335&0.455&0.760&1.335&8.165&2.830&0.440& 0.760 & 1.530 \\
        Pop & 17.515 & 6.655 & 1.305 & 2.175 & 3.865 & 21.920 & 8.270 & 1.295 & 2.240 & 4.425 \\
GPT2 Fine-tuning & \underline{26.622} & \underline{10.896} & \underline{1.992} & \underline{3.001} & \underline{14.296} & \underline{30.793} & \underline{12.799} & \underline{1.896} & \underline{2.999} & \underline{15.204} \\
Llama3 Fine-tuning & \textbf{34.008} & \textbf{17.193}  & \textbf{2.793}  & \textbf{4.202}  & \textbf{18.693}  & \textbf{38.294}  & \textbf{19.399}  & \textbf{2.393}  & \textbf{3.792}  & \textbf{19.709}  \\
        \bottomrule
    \end{tabular}
    }
\end{table*}

\begin{table*}[t]
    \centering
    \caption{Diversity and novelty performance comparison $(\%)$}
    \resizebox{\textwidth}{!}{
    \begin{tabular}{lccccccccc cccccccccc cccccccccc cccccccccc} \toprule
        & \multicolumn{3}{c}{K = 5} & \multicolumn{3}{c}{K = 10} & \multicolumn{3}{c}{K = 15} & \multicolumn{3}{c}{K = 20} \\
        \cmidrule(lr){2-4} \cmidrule(lr){5-7} \cmidrule(lr){8-10} \cmidrule(lr){11-13}
        Model & Coverage@5 $\uparrow$ & Gini Index@5 $\uparrow$ & EPC@5 $\uparrow$ & Coverage@10 $\uparrow$ & Gini Index@10 $\uparrow$ & EPC@10 $\uparrow$ & Coverage@15 $\uparrow$ & Gini Index@15 $\uparrow$ & EPC@15 $\uparrow$ & Coverage@20 $\uparrow$ & Gini Index@20 $\uparrow$ & EPC@20 $\uparrow$\\
        \midrule

       Pop & 0.160 & 80.000 & 3.410 & 0.320 & 90.000 & 4.505 &0.480&93.330&4.800& 0.640 & 95.000 & 5.180 \\
 PMF&  0.220&80.455&0.945&0.395&90.175&1.205 &0.585&93.525&1.350&0.760&95.095&	1.335 \\
        NMF &  0.220&80.455&0.945 &0.395&90.175&1.205 & 0.585&93.525&1.350 &0.760&95.095&	1.335 \\
GPT4-turbo zero-shot&1.140&96.050& 0.050&1.910&97.860 &0.090 &1.910&97.860&0.090 &2.890 & 98.740& 0.130  \\

        Item-based KNN & 6.035&96.375&0.305 & 9.675&97.775&0.400&12.485 &98.375 &0.690 &12.485&	98.375&	0.690 \\
        User-based KNN &6.270&96.850&3.435& 9.295&98.175&1.650&12.005&98.735&1.825& 14.280& 	99.020& 	1.955 \\
    GPT4-turbo few-shot   &14.872&99.771&0.250 &27.540&99.871&0.250&37.740&\underline{99.900}&0.250& 47.010&	99.900&	0.360 \\
 GPT4o  zero-shot & 17.375 & 97.720 & 0.115 & 32.910& 98.640 & 0.170 &48.810&98.640&0.170&65.090& 99.230 & 0.200   \\

Llama3 Fine-tuning & 22.505 & 99.696 & \textbf{14.399} & 31.006 & 99.692&  \textbf{15.991} & 36.735 & 99.832 &  \textbf{16.710} & 39.802 & 99.807& \textbf{17.198} \\
GPT2 Fine-tuning&23.004&95.896 & \underline{10.393} & 25.307 & 95.699 & \underline{11.301} & 25.894 & 96.004 & \underline{11.798} & 26.092 &96.505&\underline{12.291} \\

  GPT4o  few-shot& \underline{32.400}&\underline{99.890}&0.230& \underline{53.750}& \underline{99.890}&0.230&\underline{68.960}&99.940&0.230&\underline{79.030}&	\underline{99.950}&0.230\\
        Random  &  \textbf{54.160} & \textbf{99.930} &0.090 & \textbf{78.795} & \textbf{99.950} & 0.150 & \textbf{90.580}&\textbf{99.955}&0.265&\textbf{95.900}  &\textbf{99.960}  &0.195 \\
        \bottomrule
    \end{tabular}
    }
\end{table*}

\begin{table*}[tbp]
    \centering
       \caption{Performance comparison ($\%$) on cold start scenario}
    \resizebox{\textwidth}{!}{
    \begin{tabular}{lcccccccccccccccc} \toprule
        & \multicolumn{8}{c}{K = 5} \\
        \cmidrule(lr){2-9}
        Model & Hit Ratio@5 $\uparrow$ & Recall@5 $\uparrow$ & Precision@5 $\uparrow$ & F1@5 $\uparrow$ & nDCG@5 $\uparrow$ & Coverage@5 $\uparrow$ & Gini Index@5 $\uparrow$ & EPC@5 $\uparrow$ \\
        \midrule
        Random &0.000&0.000&0.000&0.000&0.000&\textbf{53.080}&\textbf{99.930}&0.000 \\
         PMF &0.000&0.000&0.000&0.000&0.000&0.190&80.190&0.000 \\
        NMF &0.000&0.000&0.000&0.000&0.000&0.190&80.190&0.000 \\
        GPT4-turbo&0.200&0.200&0.040&0.070&0.090&\underline{33.290}&\underline{98.710}&0.050  \\
        Item-based KNN &0.430&	0.430&0.090&0.140&0.220&11.660&97.280&0.150 \\
        User-based KNN &0.430&0.430&0.090&0.140&0.430&11.280&98.650&0.430 \\
        GPT4o&1.080&1.080&0.220&0.360&0.820&23.630&\underline{98.710}&0.740\\
        Pop &\underline{4.730}&\underline{4.730}&\underline{0.950}&\underline{1.580}&\underline{3.030}&0.160&80.000&\underline{2.470}\\
        GPT2 Fine-tuning & 4.409 & 4.409 & 0.882 & 1.470 & 2.540 & 9.037 & 83.844 & 1.920 \\
        Llama3 Fine-tuning & \textbf{13.613} & \textbf{13.613} & \textbf{2.723} & \textbf{4.538} & \textbf{11.810} & 7.795 & 83.029 & \textbf{11.473} \\
        \midrule
        & \multicolumn{8}{c}{K = 10} \\
        \cmidrule(lr){2-9}
        Model & Hit Ratio@10 $\uparrow$ & Recall@10 $\uparrow$ & Precision@10 $\uparrow$ & F1@10 $\uparrow$ & nDCG@10 $\uparrow$ & Coverage@10 $\uparrow$ & Gini Index@10 $\uparrow$ & EPC@10 $\uparrow$ \\
        \midrule
        Random &0.220&0.220&0.020&0.040&0.060&\textbf{77.570}&\textbf{99.950}&0.020\\
        PMF & 0.000& 0.000& 0.000& 	0.000& 0.000& 0.350& 90.100& 0.000\\
        NMF& 0.000& 0.000& 0.000& 	0.000& 0.000& 0.350& 90.100& 0.000\\
        GPT4-turbo &0.200&0.200&0.200&	0.400&0.090&\underline{61.750}&\underline{99.240}&	0.050 \\
        
        User-based KNN &1.080&1.080&0.110&0.200&0.650&15.410&99.130&0.520 \\
        Item-based KNN& 1.720&1.720&0.170&0.310&0.630&18.460&98.310&0.310\\

	GPT4o &1.510&1.510&0.150&0.270&0.960&41.520&\underline{99.240}&0.800\\	
        Pop& 6.670&6.670&	0.670&1.210&\underline{3.640}&0.320&90.000&\underline{2.710}\\
        GPT-2 Fine-tuning & \underline{7.214} & \underline{7.214} & \underline{0.721} & \underline{1.312} & 3.420 & 9.156 & 90.953  & 2.266 \\
        Llama3 Fine-tuning & \textbf{16.515} & \textbf{16.515} & \textbf{1.651} & \textbf{3.003} & \textbf{12.632} & 7.795 & 90.631 & \textbf{11.904} \\
        \bottomrule
    \end{tabular}
    }
\end{table*}

\subsection{Implementation Details} 

We first split the dataset based on user history, using 80\% of the user interaction data for training and 20\% for testing. After processing, we randomly sampled 1000 records from the test set for evaluation due to token limitations and expensive costs. For each user, we input their previously interacted items in order and use the LLM to recommend a list of course IDs they might interact with next.

For the pre-trained models, we access GPT4 turbo and GPT4o using OpenAI's API. For the fine-tuned models, we utilize Llama-3-8B and GPT-2-1.5B. The Llama-3 model is tokenized using Autotokenizer, while the GPT-2 model is tokenized using the GPT-2tokenizer.

\section{Results}

\subsection{Recommendation Performance}

To assess the recommendation capability of large language models (LLMs), we conducted experiments comparing pre-trained and fine-tuned LLMs with traditional models. The results are presented in Table 1.

In summary, we found that the performance of LLMs in the zero-shot prompting setup was relatively low compared to baseline models, outperforming only random recommendation approaches. In contrast, the few-shot prompting setup generally yielded better results, suggesting that providing historical enrollment data helps LLMs identify enrollment patterns and improve recommendation accuracy. However, overall, pre-trained LLMs still fall short of traditional recommendation methods. Two key factors may explain this outcome. First, in our prompt design, we represent each user's courses solely by their IDs to mitigate hallucination issues. Although we uploaded the course information file as a knowledge source, incorporating it through file search presents challenges. This may restrict LLMs' ability to capture semantic nuances, which are crucial for addressing cold-start problems. Additionally, due to prompt length limitations, it is not feasible to include the entire course information set in the prompt for each user. As a result, relying solely on LLMs for sequential recommendation tasks may not be optimal. Further research is needed to integrate additional guidance and constraints to help LLMs accurately capture historical user interests and produce meaningful recommendations. However, we observed that fine-tuned LLMs outperformed all other methods across various K values. Fine-tuning allows LLMs to adapt specifically to the recommendation task, enabling them to better capture user behavior and course relationships, leading to more accurate predictions.

\subsection{Diversity and Novelty Performance}

We also aim to assess the extent of diversity and novelty in the recommendations generated by LLMs, based on the results presented in Table 2.

Overall, the Random model achieves the highest Coverage and Gini Index across all K values, outperforming all other models. This result is not surprising, given its random nature, which ensures a broad range of items are recommended. In contrast, LLMs perform relatively better in diversity and novelty dimensions. Notably, the advanced model, GPT4o, outperforms GPT4-turbo across all dimensions. Upon closer examination of the generated recommendations, we observed that GPT4-turbo often exhibits “lazy behaviors”, generating similar or repetitive recommendation lists, which leads to low diversity. Furthermore, as observed in accuracy performance, few-shot models consistently outperform zero-shot models. By learning from user-specific data, few-shot models can generate more personalized and relevant recommendations, enhancing both diversity and novelty. Finally, fine-tuned models like GPT-2 and Llama3 not only excel in diversity but also significantly surpass other models in novelty. This indicates that directly applying LLMs to recommendation tasks is challenging because the data used for pre-training LLMs differs significantly from the specific requirements of recommendation tasks. However, fine-tuned LLMs, when trained on specific data, can deliver better results.

\subsection{Cold Start Scenario}

Cold start is a well-known challenge in course recommendation systems, especially in MOOC environments. It refers to the difficulty of recommending relevant courses to new users who lack sufficient interaction data. To investigate the performance of LLMs in cold start scenarios for course recommendations, we adopt a two-step approach inspired by previous studies \cite{di2023evaluating,dai2023uncovering}. First, we identify cold-start users by dividing the users of the dataset into quartiles based on their historical interaction data. The lower quartile, representing users with the least interaction, is selected as the subset of cold-start users. This method allows us to evaluate the models under consistent cold-start conditions, ensuring that all models are tested with a similar subset of users (note that we fine-tuned our LLMs using the same training set in this experiment). The results of this evaluation are presented in Table 3.

We observe that off-the-shelf LLMs outperform traditional models for cold start scenarios when only limited training data is available. Notably, LLMs do not require extensive training data to function as recommendation systems, as their pre-trained knowledge allows them to make informed predictions. The Pop method, by contrast, performs well because it simply recommends the most popular courses in the dataset. Moreover, fine-tuned LLMs achieve the highest values across all dimensions, even with minimal training data. This demonstrates that the reasoning capabilities and vast knowledge embedded in LLMs enable them to generate better recommendations. 

Secondly, we investigate the amount of training data required for traditional recommendation models to achieve performance comparable to or better than LLMs. Specifically, we chose the User-based KNN model as it performed well in the first experiment. We then evaluated their performance after training on varying proportions of training data and compared their performance to that of LLMs. Recall@5 and nDCG@5 are reported in Figure 2. As expected, the performance of User-based KNN improves with increasing amounts of training data. Also, we can observe that although GPT4-turbo's performance is not good, direct use of GPT4o as a recommendation system without training data still outperforms User-based KNN that trained on few data, i.e., less than 30\%.

Based on these findings, we conclude that using LLMs as course recommendation systems is a promising approach for mitigating the cold-start problem, offering effective solutions when traditional methods may struggle.

\begin{figure}[tb]
\centering
\includegraphics[scale=0.32]{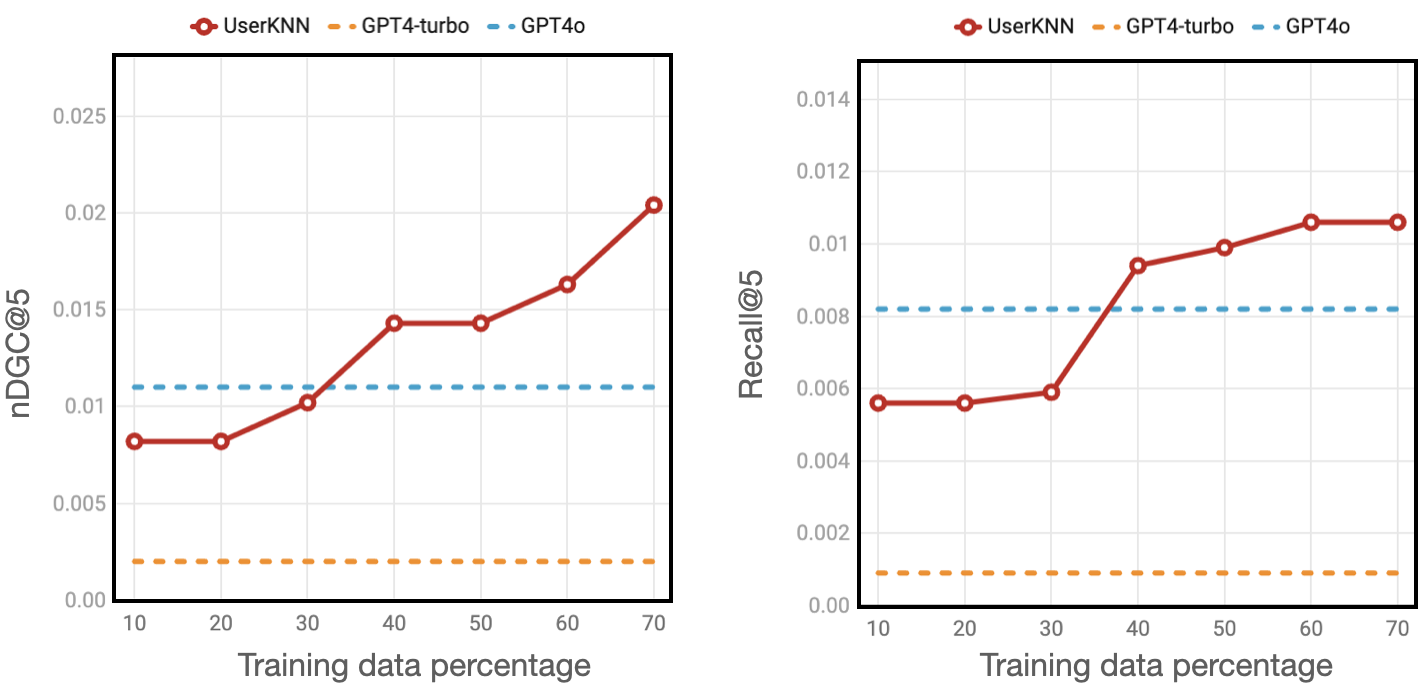}
\caption{Comparison with UserKNN in terms of different percentages of training data.}
\end{figure}




\section{Disscussion and Conclusions}

In this paper, we evaluate the performance of large language models (LLMs) in course recommendation tasks and compare them with traditional recommendation models across various dimensions and scenarios. The experimental results reveal that directly using LLMs in sequential recommendation tasks results in relatively poor performance, indicating the need for further exploration and refinement in this area. However, fine-tuned LLMs perform exceptionally well, surpassing traditional recommendation models and demonstrating promising results in cold-start scenarios. Our preliminary results provide valuable insights into the strengths and limitations of LLMs in course recommendation tasks, highlighting their potential and current challenges. We hope that our findings will inspire future research focused on enhancing course recommendation systems through the use of large language models. 

This work has some limitations. First, the experiments are conducted on a single MOOC dataset. While the results provide valuable insights, they may not generalize across different educational platforms, course types, or demographic groups. Future work can include testing the models on other datasets from university environments to evaluate their generalizability. Moreover, we used only the course IDs to represent each user's courses in our prompt design. Although this could reduce the potential for hallucinations and we provided a separate file containing course information as a knowledge source, this approach may limit the LLMs' capacity to fully capture the semantic details of the courses. 

For future work, we plan to explore better methods of incorporating user interaction data into LLMs, as this could significantly improve their ability to make personalized recommendations. Additionally, we aim to test these models on larger and more diverse datasets to further assess their generalizability. Another promising direction for future work lies in leveraging the reasoning capabilities of LLMs to offer explanations for the recommendations they generate. Providing students or system designers with clear, understandable explanations for why specific courses are recommended could enhance the transparency, persuasiveness, and trustworthiness of the recommendation system, thereby improving user satisfaction and guiding their learning goals \cite{ma2021courseq, ma2024survey}. To evaluate the potential of LLMs in this area, we want to use LLMs to generate explanations that justify a user’s preference towards recommended courses, helping to bridge the gap of recommendation performance, system interactivity, and explainability.


%
\bibliographystyle{abbrv}
\bibliography{sigproc} 
%

\balancecolumns
\end{document}